\documentclass[letterpaper,twocolumn,10pt]{article}
\usepackage{usenix,epsfig,endnotes, multirow}
\usepackage[hyphens]{url}
\usepackage{cite}
\usepackage{pbox}
\usepackage{graphicx}
\usepackage{parskip}
\usepackage{titlesec}
\usepackage{lipsum}
\usepackage{caption}
\usepackage{subcaption}

\titlespacing{\section}{1pt}{\parskip}{-\parskip}
\titlespacing{\subsection}{1pt}{\parskip}{-\parskip}
\titlespacing{\subsubsection}{1pt}{\parskip}{-\parskip}

\newenvironment{packed_enum}{
\begin{enumerate}
  \setlength{\itemsep}{1pt}
  \setlength{\parskip}{0pt}
  \setlength{\parsep}{0pt}
}{\end{enumerate}}

\newenvironment{packed_item}{
\begin{itemize}
  \setlength{\itemsep}{1pt}
  \setlength{\parskip}{0pt}
  \setlength{\parsep}{0pt}
}{\end{itemize}}

\begin{document}

\date{}

\title{Android Permissions Remystified:\\A Field Study on Contextual Integrity}

\author{
{\rm Primal Wijesekera$^1$, Arjun Baokar$^2$, Ashkan Hosseini$^2$, Serge Egelman$^2$,}\\
{\rm David Wagner$^2$, and Konstantin Beznosov$^1$}\\
$^1$University of British Columbia, Vancouver, Canada,\\
\vspace{0.5em}
\{primal,beznosov\}@ece.ubc.ca\\
$^2$University of California, Berkeley, Berkeley, USA,\\
\{arjunbaokar,ashkan\}@berkeley.edu, \{egelman,daw\}@cs.berkeley.edu\\
} 

\maketitle


\subsection*{Abstract}
Due to the amount of data that smartphone applications can potentially access, platforms enforce permission systems that allow users to regulate how applications access protected resources. If users are asked to make security decisions too frequently and in benign situations, they may become habituated and approve all future requests without regard for the consequences. If they are asked to make too few security decisions, they may become concerned that the platform is revealing too much sensitive information. To explore this tradeoff, we instrumented the Android platform to collect data regarding how often and under what circumstances smartphone applications are accessing protected resources regulated by permissions. We performed a 36-person field study to explore the notion of ``contextual integrity,'' that is, how often are applications accessing protected resources when users are not expecting it? Based on our collection of 27 million data points and exit interviews with participants, we examine the situations in which users would like the ability to deny applications access to protected resources. We found out that at least 80\% of our participants would have preferred to prevent at least one permission request, and overall, they thought that over a third of requests were invasive and desired a mechanism to block them.

\section{Introduction}

Mobile platforms enforce permission models to regulate how applications access certain resources, such as users' personal information or sensor data (e.g., camera, GPS, etc.). For instance, Android prompts the user during application installation with a list of all the abilities that that application may use in the future; if the user is uncomfortable granting any of these requests, her only option is to discontinue installing the application~\cite{AndroidPermissions2}. On iOS, the user is prompted at runtime when an application requests any one of a handful of data types for the first time, such as location, address book contacts, or photos~\cite{OGrady2014}.

Research has shown that few people read the Android permission requests and even fewer comprehend them~\cite{Felt2012}. Another problem is habituation: on average, Android applications present the user with four permission requests during the installation process~\cite{Felt2011b}. While iOS users are likely to see far fewer permission requests than Android users, because there are fewer possible permissions and they are only displayed the first time the data is actually requested, it is not clear whether or not users are being prompted about access to data that they actually find concerning, or whether they would approve of subsequent requests~\cite{Felt2012c}.

Nissenbaum posited that the reason why most privacy models fail to predict violations is that they fail to consider contextual integrity~\cite{Nissenbaum2004}. That is, privacy violations occur when personal information is used in ways that defy users' expectations. We believe that this notion of ``privacy as contextual integrity'' can be applied to smartphone permission systems to yield more effective permissions by only prompting users when an application's access to sensitive data is likely to defy expectations. As a first step down this path, we examined how applications are currently accessing this data and then examined whether or not it complied with users' expectations.

We modified Android to log whenever an application accessed a resource that was protected by application permissions and then gave these modified smartphones to 36 participants who used them as their primary phones for one week. The purpose of this was to perform dynamic analysis to determine how often various applications are actually accessing protected resources under realistic circumstances. Afterwards, subjects returned to the laboratory to return the phones and complete exit surveys. We showed them various instances over the past week where applications had accessed certain types of data and asked whether those instances were expected, and whether they would have denied access, if given the opportunity. Participants stated a desire to block a third of the requests, and that their decision processes were governed by two related underlying factors: whether they had privacy concerns surrounding the specific data type and whether they understood why the application needed it.

We contribute the following:
\begin{packed_item}
\item To the best of our knowledge, we performed the first field study to quantify the permission usage by third party applications under realistic circumstances.
\item We show that our participants wanted to block access to protected resources a third of the time. This suggests that some requests should be granted by runtime consent dialogs, rather than the current all-or-nothing install-time approval approach.
\item We model participants' decisions and show how a runtime classifier may be able to determine when to confront users with permission decisions.
\end{packed_item}

\section{Related Work}

While users are required to approve Android application permission requests prior to installation, most users do not pay attention to these requests, and fewer fully comprehend them~\cite{Felt2012,Kelley2012}. In fact, studies have shown that even developers are not fully knowledgeable about permissions~\cite{Stevens2013}, and are given a lot of freedom when posting an application to the Google Play Store~\cite{Barrera2012}. Applications often do not follow the principle of least privilege, intentionally or unintentionally~\cite{Wei2012}. Other work has made suggestions on improving the Android permission model with better definitions and hierarchical breakdowns~\cite{Barrera2010}. Some researchers have experimented with adding fine-grained access control to the Android model~\cite{Bugiel2013}. Providing users with more privacy information and personal examples has been shown to help users in choosing applications with fewer permissions~\cite{Kelley2013, Harbach2014}. 

Previous work has examined the overuse of permissions by applications~\cite{Felt2011b, Gorla2014}, and attempted to identify malicious applications through their permission requests~\cite{Sarma2012}, or through natural language processing of application descriptions~\cite{Pandita2013}. Researchers have additionally developed static analysis tools to analyze Android permission specifications~\cite{Felt2011b, Au2012, Bodden2013}. Felt et al.\ created a permission map through static analysis of many Android applications. They found that roughly one-third of their tested applications were over-privileged. Our work complements this static analysis by applying dynamic analysis to permission usage. Other dynamic analysis has been applied to native (non-Java) APIs among third-party mobile markets~\cite{Spreitzenbarth2013}, whereas we apply it to the Java APIs available to developers in the Google Play Store.

Researchers examined user privacy expectations surrounding application permissions, and found that users were often surprised by the abilities of background applications to collect data~\cite{Jung2012, Thompson2013}. Their level of concern varied from annoyance to seeking retribution when presented with possible risks associated with permissions~\cite{Felt2012c}. Some studies employed crowdsourcing to create a privacy model based on user expectations~\cite{Lin2012}.

Researchers have designed systems to track or reduce privacy violations by recommending applications based on users' security concerns~\cite{Enck2010, Hornyack2011, Zhu2014, Gibler2012, Klieber2014, Xu2012, Almohri2014, Zhang2013}. Other tools dynamically block runtime permission requests~\cite{Shebaro2014}. Enck et al.\ found that a considerable number of applications transmitted location or other user data to third parties without requiring user consent~\cite{Enck2010}. Hornyack et al.'s AppFence system gave users the ability to deny data to applications or substitute fake data~\cite{Hornyack2011}. However, this broke application functionality for one-third of the applications tested.

Reducing the number of security decisions a user must make at install-time or run-time is likely to decrease habituation, and therefore, it is critical to identify {\it which} security decisions users should be asked to make. Based on this theory, Felt et al.\ created a decision tree to aid platform designers in determining the most appropriate permission-granting mechanism for a given resource (e.g., access to benign resources should be granted automatically, whereas access to dangerous resources should require approval)~\cite{Felt2012b}. They concluded that the majority of Android permissions can be automatically granted, but 16\% (corresponding to the 12 permissions in Table \ref{tbl:perm-list}) should be granted via runtime dialogs.

Nissenbaum's theory of contextual integrity can help us to analyze ``the appropriateness of a flow" in the context of permissions granted to Android applications~\cite{Nissenbaum2004}. There is ambiguity in defining when an application actually needs access to user data to run properly. It is quite easy to see why a location-sharing application would need access to GPS data, whereas that same request coming from a game like Angry Birds is less obvious. ``Contextual integrity is preserved if information flows according to contextual norms''~\cite{Nissenbaum2004}, however, the lack of thorough documentation on the Android permission model makes it easier for programmers to neglect these norms, whether intentionally or accidentally~\cite{Shklovski2014}. Deciding on whether an application is violating users' privacy can be quite complicated since ``the scope of privacy is wide-ranging''~\cite{Nissenbaum2004}. To that end, we performed dynamic analysis to measure how often (and under what circumstances) applications were accessing protected resources, whether this complied with users' expectations, as well as how often they might be prompted if we adopt Felt et al.'s proposal to require runtime user confirmation before accessing a subset of these resources~\cite{Felt2012b}.

\section{Methodology}

Our long-term research goal is to minimize habituation by only confronting users with {\it necessary} security decisions by not showing them permission requests that are either expected, reversible, or unconcerning. In this study, we explored the problem space in two parts: we instrumented Android so that we could collect actual usage data to understand how often access to various protected resources is requested by applications in practice, and then we surveyed our participants to understand the requests that they would not have granted, if given the option. This field study involved 36 participants over the course of one week of normal smartphone usage. In this section, we describe the log data that we collected, our recruitment procedure, and then our exit survey.



\subsection{Tracking Access to Sensitive Data}

In Android, when applications attempt to access protected resources (e.g., personal information, sensor data, etc.), the operating system checks to see whether or not the requesting application has been granted permission. We modified the Android platform to add a logging framework so that we could determine every time one of these resources was accessed by an application. Because our target device was a Samsung Nexus S smartphone, we modified Android 4.1.1 (Jellybean), which was the newest version of Android supported by our hardware.

\subsubsection{Data Collection Architecture}

Our goal was to collect as much data as possible surrounding each applications' access to protected resources, while minimizing our impact on system performance. Our data collection framework consisted of two main components: a series of ``producers'' that 
hooked various Android API calls and a ``consumer'' embedded in the main Android framework service that wrote the data to a log file and uploaded it to our collection server. 


We logged three kinds of permission requests. First, we logged function calls checked by {\tt checkPermission()} in the Android {\tt Context} implementation. Instrumenting the {\tt Context} implementation, instead of the {\tt ActivityManagerService} or {\tt PackageManager}, allowed us to also log the function name invoked by the user-space application. Next, we logged access to the {\tt ContentProvider} class, which verifies the read and write permissions of an application prior to it accessing structured data (e.g., contacts or calendars)~\cite{AndroidContentProviders}. Finally, we tracked permission checks during {\tt Intent} transmission, by instrumenting the {\tt ActivityManagerService} and {\tt BroadcastQueue}. {\tt Intents} allow an application to pass messages to another application when an activity is to be performed in that other application (e.g., opening a URL in the web browser)~\cite{AndroidIntents}.

We created a component called {\tt Producer}, which fetches the data from the above instrumented points and sends it back to the {\tt Consumer}, which is responsible for logging everything reported. {\tt Producers} are scattered across the Android Platform, since permission checks occur in multiple places. We placed the {\tt Producer} that fetched the most data in {\tt system\_server}, which recorded direct function calls to Android's Java API. For a majority of privileged function calls, when a user application invokes the function, it sends the request to {\tt system\_server} via {\tt Binder}. {\tt Binder} is the most prominent IPC mechanism implemented to communicate with the Android Platform (whereas {\tt Intents} communicate between applications). For requests that do not make IPC calls to the {\tt system\_server}, a {\tt Producer} is placed in the user application context (e.g., in the case of {\tt ContentProviders}).




The {\tt Consumer} class is responsible for logging data produced by each {\tt Producer}. Apart from this, the {\tt Consumer} also stores contextual information alongside the permission details. More details on the contextual data is given in Section \ref{sec:collection}. The {\tt Consumer} syncs data with the filesystem periodically to minimize impact on system performance. All log data is written to the internal storage of the Android Phone because the Android kernel is not allowed to write to external storage for security reasons. Although this protects our data from curious or careless users, it also limits our storage capacity. Thus, we compressed the log files once every two hours and upload them to our collection servers whenever the phone had an active Internet connection (the average uploaded and zipped log file was around 108KB and contained 9,000 events).


Due to the high volume of permission checks we encountered and our goal of keeping system performance at acceptable levels, we added logic so that the {\tt Consumer} can rate-limit itself. Specifically, if it has logged permission checks for a particular application/permission combination more than 10,000 times, it examines whether it did so while exceeding an average rate of 1 permission check every 2 seconds. If so, the {\tt Consumer} will only record 10\% of all future requests for this application/permission combination. When this rate-limiting is enabled, the {\tt Consumer} tracks these application/permission combinations and updates all the {\tt Producers} so that they start dropping these log entries. Finally, the {\tt Consumer} makes a note of whenever this occurs so that we can extrapolate the true number of permission checks that occurred.

\subsubsection{Data Collection}
\label{sec:collection}

We hooked the permission-checking APIs so that every time system code was called to check whether an application had been granted a particular permission, we logged the name of the permission being checked, the name of the calling application, the API method that resulted in the permission check, and various contextual data. Recall that permission checks not only occur when certain protected API methods are called, but also when protected {\tt Intents} and {\tt ContentProviders} are accessed. Thus, our logs differentiate between these three ways of accessing protected resources.


With regard to contextual data, in addition to timestamps, we collected the following types of data:

\begin{packed_item}
\item {\bf Visibility}---Whether the requesting application was visible to the user or not, which we categorized into four sub-categories: the application was running (a) as a service with no visibility to the user; (b) as a service, but interacted with the user via notifications or sounds; (c) as a foreground process, but was in the background due to multitasking; or (d) as a foreground process with which the user was interacting.
\item {\bf Screen Status}---Whether the screen was on/off.
\item {\bf Connectivity}---Whether the phone was connected to a WiFi network.
\item {\bf Location}---The user's last known coordinates. In order to preserve battery life, we collected cached location data, rather than directly querying the GPS.
\item {\bf View}---The UI elements in the requesting application that were exposed to the user at the time that a protected resource was accessed. Specifically, since the UI is built from an XML file, we recorded the name of the screen as defined in the DOM.
\item {\bf History}---A list of applications with which the user interacted prior to the requesting application.
\item {\bf Path}---When access to a {\tt ContentProvider} object was requested, the path to the specific content (e.g., photos, contacts, etc.).
\end{packed_item}


Felt et al.\ proposed that most Android permissions should require no {\it a priori} user approval, but 12 permissions (Table \ref{tbl:perm-list}) should be granted at runtime so that users have contextual information to infer why the data might be needed~\cite{Felt2012b}. Specifically, if the user is asked to grant a permission while using an application, she may have some understanding of why the application needs that permission based on what she was doing. We initially wanted to perform experience sampling by probabilistically questioning participants whenever any of these 12 permissions were checked~\cite{Larson1983}. Our goal was to survey participants about whether access to these resources was expected and whether it should proceed, but we were concerned that this would prime them to the security focus of our experiment, biasing their subsequent behaviors. Instead, we instrumented the phones to probabilistically take screenshots of what participants were doing when these 12 permissions were checked so that we could ask them about it during the exit survey. We used reservoir sampling to minimize storage and performance impacts, while also ensuring that the screenshots covered a broad set of applications and permissions~\cite{Vitter1985}.

\begin{table}[t]
\small
\center
\begin{tabular}{|l|p{4.3cm}|}
\hline
\textbf{Permission Type} & \textbf{Activity} \\ \hline
\begin{tabular}[c]{@{}l@{}}WRITE\_SYNC\_\\ SETTINGS\end{tabular} & Change sync settings for an application when the user is roaming \\ \hline
\begin{tabular}[c]{@{}l@{}}ACCESS\_WIFI\_\\ STATE\end{tabular} & View nearby SSIDs \\ \hline
INTERNET & Access Internet when the user is roaming \\ \hline
NFC & Communicate via NFC \\ \hline
\begin{tabular}[c]{@{}l@{}}READ\_HISTORY\_\\ BOOKMARKS\end{tabular} & Read users' browser history \\ \hline
\begin{tabular}[c]{@{}l@{}}ACCESS\_FINE\_\\ LOCATION\end{tabular} & Read the GPS location \\ \hline
\begin{tabular}[c]{@{}l@{}}ACCESS\_COARSE\_\\ LOCATION\end{tabular} & Read the network-inferred location (i.e., cell tower and/or WiFi) \\ \hline
\begin{tabular}[c]{@{}l@{}}LOCATION\_\\ HARDWARE\end{tabular} & Directly access GPS data \\ \hline
READ\_CALL\_LOG & Read call history \\ \hline
ADD\_VOICEMAIL & Read call history \\ \hline
READ\_SMS & Read sent/received/draft SMS \\ \hline
SEND\_SMS & Send SMS \\ \hline
\end{tabular}
\begin{flushleft}
\caption{The 12 permissions that Felt et al.\ recommend be granted via runtime dialogs~\cite{Felt2012b}. We randomly took screenshots when these permissions were requested by applications, and we asked about them in our exit survey.}
\label{tbl:perm-list}
\end{flushleft}
\end{table}

\begin{figure}[t]
\begin{subfigure}[h]{0.5\textwidth}
\centering
\includegraphics[scale=0.2]{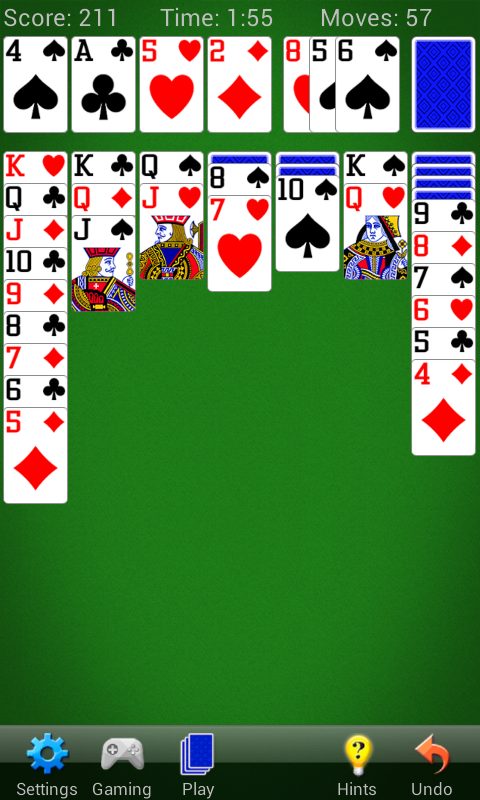}
\caption{Screenshot}
\label{fig_scrnshot}
\end{subfigure}
\begin{subfigure}[h]{0.5\textwidth}
\centering
\small
\begin{tabular}{|l|p{5cm}|}
\hline
\textbf{Name} & \textbf{Log Data} \\ \hline
Type & API\_FUNC \\ \hline
Permission & \pbox{20cm}{ACCESS\_WIFI\_STATE} \\ \hline
App\_Name & com.spotify.music \\ \hline
Timestamp & 1412888326273 \\ \hline
API Function & getScanResults() \\ \hline
Visibility & FALSE \\ \hline
Screen Status & SCREEN\_ON \\ \hline
Connectivity & NOT\_CONNECTED \\ \hline
Location & Lat 37.8735436 Long -122.2992491 - 1412538686641 (Time it was updated) \\ \hline
View & com.mobilityware.solitaire/.Solitaire \\ \hline
History &  \pbox{20cm}{com.android.phone/.InCallScreen \\ com.android.launcher/com.android.- \\ launcher2.Launcher \\ com.android.mms/ConversationList}   \\ \hline
Path & N/A \\ \hline
Screenshot & 898448929 \\ \hline
\end{tabular}
\caption{Corresponding log entry}
\label{tbl:smplLog}
\end{subfigure}
\caption{Screenshot (a) and corresponding log entry (b) captured during the experiment.}
\label{fig:example}
\end{figure}

Figure \ref{fig:example} shows an example screenshot captured during the study along with its corresponding
log entry. The user was playing the Solitaire game while Spotify requested a WiFi scan. Since
this function was of interest (Table \ref{tbl:perm-list}), our instrumentation took a screenshot.  Since Spotify was not the application the participant was
interacting with, its visibility is set to {\it false}. The history shows that prior to Spotify calling {\tt getScanResults()}, the user had viewed Solitaire, the call screen, the launcher, and the list of MMS conversations.

\subsection{Recruitment}

We placed an online recruitment advertisement on Craigslist in October of 2014, under the ``et cetera jobs'' 
section.\footnote{Approved by our IRB under protocol \#2013-02-4992} The title of the advertisement was ``Research Study on Android Smartphones,'' and it 
stated that the study was about how people interact with their smartphones. We made no mention of 
security or privacy. Those interested in participating were directed to an online consent form. Upon agreeing to the consent form, potential participants were directed to a screening application in the Google Play store. The screening application asked for information about each potential participant's age, 
gender, smartphone make and model. It also collected data on their phones' internal memory size and the installed applications. We screened out applicants who 
were under 18 years of age or used providers other than T-Mobile, since our experimental phones could not attain 3G speeds on other providers. We collected data on participants' installed applications so that we could pre-install their free applications on our experimental phones, prior to them visiting our laboratory. (We copied their paid applications from their phones, since we could not download those from Google Play ahead of time.)

We contacted participants who met our screening requirements by email to schedule a time for them to come and visit us to do the initial setup. Overall, 48 people showed up to our laboratory, and out of those 48 people, 40 qualified (8 had to be rejected because our screening application was unable to distinguish some Metro PCS users from T-Mobile users).  In the email we
noted that due to the space constraints of our experimental phones, we might not be able to install all the applications present on their existing phones, and therefore they needed to make a note of the ones that they planned to use during the next week. The initial setup took roughly 30 minutes and involved installing their existing SIM cards into our experimental phones, helping them set up their Google and other accounts, and making sure they had all the applications needed
for the coming week. We immediately compensated each participant with a \$35 gift card for showing up at the setup session. Out of 40 people who were given phones, 2 did not return the phones, and 2 did not regularly use the phones during the study period. Of our 36 remaining participants who used the phones regularly, 19 were 
male and 17 were female; ages ranged from 20 to 63 years old (\begin{math}\mu \end{math} = 32, \begin{math} \sigma 
\end{math}= 11).

After the initial setup session, participants used the experimental phones for one week in lieu of their normal phones. They were allowed to install and uninstall 
applications, and we instructed them to use these phones as they would their normal phones. 
Our logging framework kept track of every protected resource accessed by a user-level application along with the 
previously-mentioned contextual data. Due to storage constraints on the devices, our software uploaded log files to our server every two hours. However, to preserve participants' privacy, screenshots remained on the phones during the course of the week. At the end of the week, each participant returned to our laboratory, completed an exit survey, returned the phone, and then received an additional \$100 gift card for completing the study.

\subsection{Exit Survey}

\begin{figure}[t]
\centering
\begin{subfigure}[h]{0.5\textwidth}
\includegraphics[width=3.2in]{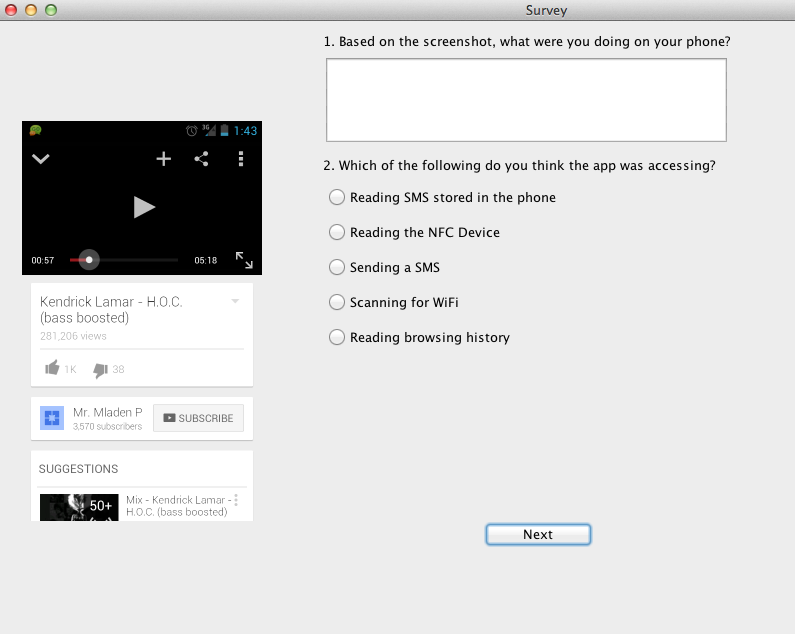}
\caption{On the first screen, participants answered questions to establish awareness of the permission request based on the screenshot.}
\label{fig:initial_screen}
\end{subfigure}
\quad
\begin{subfigure}[h]{0.5\textwidth}
\includegraphics[width=3.2in]{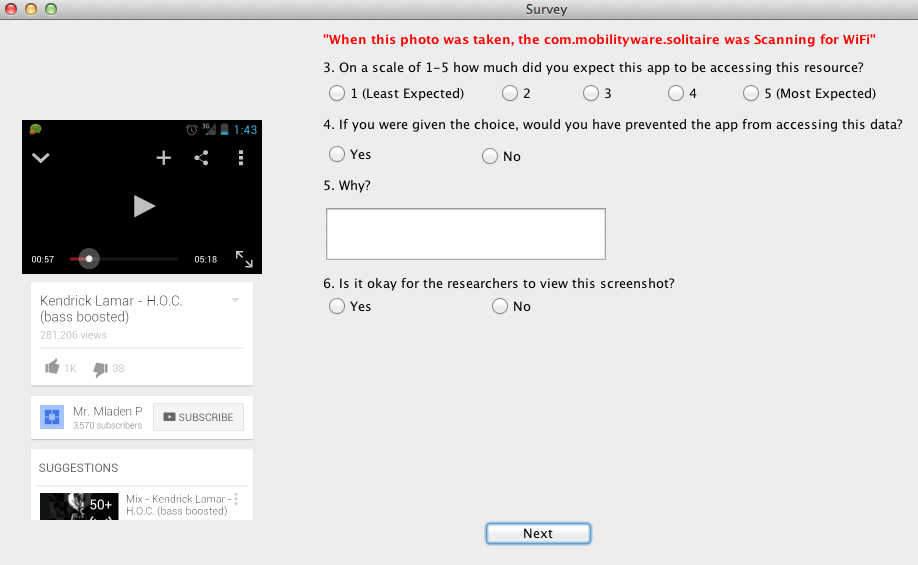}
\caption{On the second screen, they saw the resource accessed, stated whether it was expected, and whether it should have been blocked.}
\label{fig:followup_screen}
\end{subfigure}
\begin{flushleft}
\caption{Exit Survey Interface}
\label{fig:three graphs}
\end{flushleft}
\end{figure}

When participants returned to our laboratory, they completed an exit survey. The 
exit survey software ran on a laptop in a private room so that it could ask questions about what they were doing on their phones during the course of the week without raising privacy concerns. We did not view their screenshots until participants gave us permission. The survey had three components:

\begin{packed_item}
\item {\bf Screenshots}---Our software displayed a screenshot taken during the course of the week when one of the 12 resources in Table \ref{tbl:perm-list} was accessed. Next to the screenshot (Figure \ref{fig:initial_screen}), we asked participants what they were doing on the phone when the screenshot was taken (open-ended). We also asked them to indicate which of several actions they believed the application was performing, chosen from a multiple-choice list of permissions presented in plain language (e.g., ``reading browser history,'' ``sending a SMS,'' etc.). After answering these two questions, they proceeded to a second page of questions (Figure \ref{fig:followup_screen}). We informed participants at the top of this page of the resource that the application had accessed when the screenshot was taken, and asked them to indicate how much they expected this using a 5-point Likert scale. Next, we asked, ``if you were given the choice, would you have prevented the app from accessing this data,'' and to explain why or why not. Finally, we asked for permission to view that particular screenshot. This phase of the exit survey was repeated for 10-15 different screenshots per participant, based on the number of screenshots saved by our reservoir sampling algorithm.

\item {\bf Locked Screens}---The second part of our survey involved questions about the same protected resources, however, this time these resources were accessed while device screens were off (i.e., when participants were not actively using their phones). Because there were no contextual cues (i.e., screenshots), we outright told participants which applications were accessing which resources and asked them multiple choice questions about whether they wanted to prevent this and the degree to which these behaviors were expected. They answered these two questions for up to 10 different requests, similarly chosen by our reservoir sampling algorithm to yield a breadth of application/permission combinations.

\item {\bf Personal Privacy Preferences}---Finally, in order to correlate participants' responses with their personal privacy preferences, they completed two privacy scales. Because of the numerous reliability problems with the often cited Westin index~\cite{Woodruff2014}, participants completed both Buchanan et al.'s Privacy Concerns Scale (PCS)~\cite{Buchanan2007} and Malhotra et al.'s Internet Users' Information Privacy Concerns (IUIPC) scale~\cite{Malhotra04}. We compared the average scores of both scales.
\end{packed_item}

After completing the exit survey, we re-entered the room, answered any remaining questions about the experiment, and then assisted the participant in transferring her SIM card back into her personal phone. Finally, we compensated each participant with a \$100 gift card.

Three researchers independently coded 423 responses to the open-ended question from the screenshot portion of the survey. The number of responses per participant varied, as they were randomly selected based on the number of screenshots taken by their phones. Participants who used their phones more heavily had more screenshots, and thus answered more questions. Prior to meeting to achieve consensus, the three coders disagreed on the coding of 42 responses, which equated to an inter-rater agreement of 90\%. We used Fleiss' kappa to assess the reliability of ratings, taking into account the 9 possible codings for each response. The process resulted in a kappa of 0.61, which indicates substantial agreement.

\section{Application Behaviors}
\label{sec:apps}

During the week that participants used our instrumented phones, we logged 27M requests by applications to protected resources (i.e., those governed by Android permissions). This translates to over one hundred thousand requests per user per day. In this section, we quantify the circumstances under which these resources were accessed. We focus on the rate at which applications requested access to protected resources when participants were not actively using those applications (i.e., the situations that likely defy users' expectations), access to certain resources with particularly high frequency, and the impact of replacing certain requests with runtime confirmation dialogs (as per Felt et al.'s suggestion~\cite{Felt2012b}).


\subsection{Invisible Permission Requests}
\label{sec:invisible}


In many cases, it is entirely expected that an application might make frequent requests to resources protected by permissions. For instance, the INTERNET permission is used every time an application needs to open a socket, ACCESS\_FINE\_LOCATION is used every time the user's location is checked by a mapping application, and so on. However, in these cases, one expects users to have certain contextual cues to help them understand that these applications are running and making these requests. Based on our log data, most requests occurred while participants were not actually interacting with those applications, nor did they have any cues to indicate that the applications were even running. When resources are accessed, applications can be in five different states, with regard to their visibility to users:

\begin{packed_enum}
\item {\bf Visible foreground application (12.04\%)}: the user is using the application requesting the resource.
\item {\bf Invisible background application (0.70\%)}: due to multitasking, the application is in the background.
\item {\bf Visible background service (12.86\%)}: the application is a background service, but the user may be aware of its presence due to other cues (e.g., it is playing music or is present in the notification bar).
\item {\bf Invisible background service (14.40\%)}: the application is a background service without visibility.
\item {\bf Screen off (60.00\%)}: the application is running, but the phone screen is off because it is not in use.
\end{packed_enum}

Combining the 3.3M (12.04\% of 27M) requests that were granted when the user was actively using the application (Category 1) with the 3.5M (12.86\% of 27M) requests that were granted when the user had other contextual cues to indicate that the application was running (Category 3), we can see that fewer than one quarter of all permission requests (24.9\% of 27M) occurred when the user had clear indications that those applications were running. This suggests that during the vast majority of the time, access to protected resources occurs opaquely to users. We focus on these 20.3M ``invisible'' requests (75.1\% of 27M) in the remainder of this subsection.

Harbach et al.\ found that users' phone screens are off 94\% of the time on average~\cite{Harbach2014b}. We observed that 60\% of permission requests occurred while participants' phone screens were off, which suggests that permission requests occurred less frequently than when participants were using their phones. At the same time, certain applications made more requests when participants were not using their phones: ``Brave Frontier Service,'' ``Microsoft Sky Drive,'' and ``Tile game by UMoni.'' Our study collected data on over 300 applications, and therefore it is possible that with a larger sample size, we would observe other applications engaging in this behavior. All of the aforementioned applications primarily requested ACCESS\_WIFI\_STATE and INTERNET. While a definitive explanation for this behavior requires examining source code or the call stacks of these applications, we hypothesize that they were continuously updating local data from remote servers. For instance, Sky Drive may have been updating documents, whereas the other two applications may have been checking the status of multiplayer games.



\begin{table}[t]
\centering
\small
\begin{tabular}{|l|r|}
\hline
{\bf Permission} &{\bf Requests} \\ \hline
ACCESS\_NETWORK\_STATE & 31,206 \\ \hline
WAKE\_LOCK & 23,816 \\ \hline
ACCESS\_FINE\_LOCATION & 5,652 \\ \hline
GET\_ACCOUNTS & 3,411 \\ \hline
ACCESS\_WIFI\_STATE & 1,826 \\ \hline
UPDATE\_DEVICE\_STATS & 1,426 \\ \hline
ACCESS\_COARSE\_LOCATION & 1,277 \\ \hline
AUTHENTICATE\_ACCOUNTS & 644 \\ \hline
READ\_SYNC\_SETTINGS & 426 \\ \hline
INTERNET & 416 \\ \hline
\end{tabular}
\begin{flushleft}
\caption{The most frequently requested permissions by applications with zero visibility to the user.}
\label{tbl:so_perm}
\end{flushleft}
\end{table}

\begin{table}[t]
\centering
\small
\begin{tabular}{|l|r|}
\hline
{\bf Application} & {\bf Requests} \\ \hline
Facebook & 36,346 \\ \hline
Google Location Reporting & 31,747 \\ \hline
Facebook Messenger & 22,008 \\ \hline
Taptu DJ & 10,662 \\ \hline
Google Maps & 5,483 \\ \hline
Google Gapps & 4,472 \\ \hline
Foursquare & 3,527 \\ \hline
Yahoo Weather & 2,659 \\ \hline
Devexpert Weather & 2,567 \\ \hline
Tile Game(Umoni) & 2,239 \\ \hline
\end{tabular}
\begin{flushleft}
\caption{The applications making the most permission requests while running invisibly to the user.}
\label{tbl:so_app}
\end{flushleft}
\vspace{-1em}
\end{table}

Table \ref{tbl:so_perm} shows the most frequently requested permissions from applications running invisibly to the user (i.e., Categories 2, 4, and 5); Table \ref{tbl:so_app} shows the applications responsible for these requests (Appendix \ref{app:app_brekdown} lists the permissions requested by these applications). We normalized the numbers to show requests per user/day. ACCESS\_NETWORK\_STATE was most frequently requested, averaging 31,206 times per user/day---roughly once every 3 seconds. This is due to applications constantly checking for Internet connectivity. However, the 5,562 requests/day to ACCESS\_FINE\_LOCATION and 1,277 requests/day to ACCESS\_COARSE\_LOCATION are more concerning, as this could enable detailed tracking of the user's movement throughout the day. Similarly, a user's location can be inferred by using ACCESS\_WIFI\_STATE to get data on nearby WiFi SSIDs.

Contextual integrity means ensuring that information flows are appropriate, as determined by the user. Thus, users need the ability to see information flows. Current mobile platforms have done some work to let the user know about location tracking. For instance, recent versions of Android allow users to see which applications have used location data recently. While attribution is a positive step towards contextual integrity, attribution is most beneficial for actions that are reversible, whereas the disclosure of location information is not something that can be undone~\cite{Felt2012b}. We observed that fewer than 1\% of location requests were made when the applications were visible to the user or resulted in the displaying of a GPS notification icon. Given that Thompson et al.\ showed that most users do not understand that applications running in the background may have the same abilities as applications running in the foreground~\cite{Thompson2013}, it is likely that in the vast majority of cases, users do not know when their locations are being disclosed.

This low visibility rate is because Android only shows a notification icon when the GPS sensor is accessed, while offering alternative ways of inferring location. In 66.1\% of applications' location requests, they directly queried the {\tt TelephonyManager}, which can be used to determine location via cellular tower information. In 33.3\% of the cases, applications requested the SSIDs of nearby WiFi networks. In the remaining 0.6\% of cases, applications accessed location information using one of three built-in location providers: GPS, network, or passive. Applications accessed the GPS location provider only 6\% of the time (which displayed a GPS notification). In the other 94\% of the time, 13\% queried the network provider (i.e., approximate location based on nearby cellular towers and WiFi SSIDs) and 81\% queried the passive location provider. The passive location provider caches prior requests made to either the GPS or network providers. Thus, across all requests for location data, the GPS notification icon appeared 0.04\% of the time.

While the alternatives to querying the GPS are less accurate, users are still surprised by their accuracy~\cite{fu2014general}. This suggests a serious violation of contextual integrity, since users likely have no idea their locations are being requested in the vast majority of cases. Thus, runtime notifications for location tracking need to be improved~\cite{fu2014field}.

Apart from these invisible location requests, we also observed applications reading stored SMS messages (125 times per user/day), reading browser history (5 times per user/day), and accessing the camera (once per user/day). Though the use of these permission does not necessarily lead to privacy violations, users have no contextual cues to understand that these requests are occurring.

\begin{table}[t]
\centering
\small
\begin{tabular}{|l|r|r|}
\hline
{\bf Application / Permission} & {\bf Peak (ms)} & {\bf Avg. (ms)} \\
\hline
\hline
com.facebook.katana & \multirow{2}{*}{213.88} & \multirow{2}{*}{956.97} \\ \cline{1-1}
ACCESS\_NETWORK\_STATE &  &  \\ \hline
\hline
com.facebook.orca & \multirow{2}{*}{334.78} & \multirow{2}{*}{1146.05} \\ \cline{1-1}
ACCESS\_NETWORK\_STATE &  &  \\ \hline
\hline
com.google.android.apps.maps & \multirow{2}{*}{247.89} & \multirow{2}{*}{624.61} \\ \cline{1-1}
ACCESS\_NETWORK\_STATE &  &  \\ \hline
\hline
com.google.process.gapps & \multirow{2}{*}{315.31} & \multirow{2}{*}{315.31} \\ \cline{1-1}
AUTHENTICATE\_ACCOUNTS &  &  \\ \hline
\hline
com.google.process.gapps & \multirow{2}{*}{898.94} & \multirow{2}{*}{1400.20} \\ \cline{1-1}
WAKE\_LOCK &  &  \\ \hline
\hline
com.google.process.location & \multirow{2}{*}{176.11} & \multirow{2}{*}{991.46} \\ \cline{1-1}
WAKE\_LOCK &  &  \\ \hline
\hline
com.google.process.location & \multirow{2}{*}{1387.26} & \multirow{2}{*}{1387.26} \\ \cline{1-1}
ACCESS\_FINE\_LOCATION &  &  \\ \hline
\hline
com.google.process.location & \multirow{2}{*}{373.41} & \multirow{2}{*}{1878.88} \\ \cline{1-1}
GET\_ACCOUNTS &  &  \\ \hline
\hline
com.google.process.location & \multirow{2}{*}{1901.91} & \multirow{2}{*}{1901.91} \\ \cline{1-1}
ACCESS\_WIFI\_STATE &  &  \\ \hline
\hline
com.king.farmheroessaga & \multirow{2}{*}{284.02} & \multirow{2}{*}{731.27} \\ \cline{1-1}
ACCESS\_NETWORK\_STATE &  &  \\ \hline
\hline
com.pandora.android & \multirow{2}{*}{541.37} & \multirow{2}{*}{541.37} \\ \cline{1-1}
ACCESS\_NETWORK\_STATE &  &  \\ \hline
\hline
com.taptu.streams & \multirow{2}{*}{1746.36} & \multirow{2}{*}{1746.36} \\ \cline{1-1}
ACCESS\_NETWORK\_STATE &  &  \\ \hline
\end{tabular}
\begin{flushleft}
\caption{The application/permission combinations that needed to be rate limited during the study. The last two columns
show the fastest interval recorded and the average of all the intervals recorded before rate-limiting.}
\label{tbl:rate-limit}
\end{flushleft}
\end{table}

\begin{table*}[t]
\center
\small
\begin{tabular}{|l||r|r||r|r||r|r|}
\hline
\textbf{Resource} & \multicolumn{2}{|c||}{\bf Visible} & \multicolumn{2}{|c||}{\bf Invisible} & \multicolumn{2}{|c|}{\bf Total}\\
& \textbf{Data Exposed} & \textbf{Requests} & \textbf{Data Exposed} & \textbf{Requests} & \textbf{Data Exposed} & \textbf{Requests}\\ \hline
Location & 758 & 2,205 & 3,881 & 8,755 & 4,639 & 10,960 \\ \hline
Read SMS data  & 378 & 486 & 72 & 125 & 450 & 611 \\ \hline
Sending SMS  & 7 & 7 & 1 & 1 & 8 & 8 \\ \hline
Browser History  & 12 & 14 & 2 & 5 & 14 & 19 \\ \hline
{\bf Total} & 1,155 & 2,712 & 3,956 & 8,886 & 5,111 & 11,598\\ \hline
\end{tabular}
\begin{flushleft}
\caption{The sensitive permission requests (per user/day) when requesting applications were visible/invisible to users. ``Data exposed'' reflects the subset of permission-protected requests that resulted in sensitive data being accessed.}
\label{tbl:privacy_breakdown}
\end{flushleft}
\end{table*}

\subsection{High Frequency Requests}

Some permission requests occurred so frequently that a few applications (i.e., Facebook, Facebook Messenger, Google Location Reporting, Google Maps, Farm Heroes Saga) had to be rate limited in our log files, so that the logs would not fill up users' remaining storage or incur performance overhead. Our software probabilistically began dropping log entries if during an application's previous 10,000 requests for a particular permission, the average interval between requests was less than 2 seconds. Table \ref{tbl:rate-limit} shows the complete list of application/permission combinations that exceeded this threshold. For instance, the most frequent requests came from Facebook requesting ACCESS\_NETWORK\_STATE with an average interval of 213.88 ms (i.e., almost 5 times per second).

With the exception of Google's applications, all rate-limited applications made excessive requests for the phone's connectivity state. Our hypothesis is that once these applications lose connectivity, they continuously poll the system until access is regained. All of these applications' use of the {\tt getActiveNetworkInfo()} method resulted in these permission checks, which returns a {\tt NetworkInfo} object. This object allows developers to determine connection state (e.g., connected, disconnected, etc.) and type (e.g., WiFi, Bluetooth, cellular, etc.). Thus, these requests do not appear to be leaking sensitive information {\it per se}, but their frequency may have an adverse effect on performance and battery life. It is possible that using the {\tt ConnectivityManager}'s {\tt NetworkCallback} method may be able to fulfill this need with far fewer permission checks.

\subsection{Feasibility of Runtime Requests}


Felt et al.\ posited that while most application permissions can be granted automatically in order to not habituate users to relatively benign risks, certain permissions requests should require runtime consent~\cite{Felt2012b}. They advocated using runtime dialogs before the following abilities, observed during our study, should proceed:

\begin{packed_enum}
\item Reading the user's location information, which includes using a conventional location API or scanning for nearby WiFi SSIDs.
\item Reading the user's web browser history.
\item Reading saved SMS messages.
\item Sending SMS messages that incur charges, or inappropriately spamming the user's contact list.
\end{packed_enum}

These four resources/abilities are governed by the 12 Android permissions listed in Table \ref{tbl:perm-list}. Of the 300 applications that we observed during the experiment, 91 (30.3\%) performed one of these abilities during the study period. On average, these permissions were requested 213 times per hour per user---roughly every 20 seconds. However, permission checks occur under a variety of circumstances, only a subset of which allow applications access to sensitive resources. As a result, platform developers may decide to only show runtime warnings to users when protected data is read or modified. Thus, we attempted to quantify the frequency with which permission checks actually result in access to sensitive resources for each of these four categories. Table \ref{tbl:privacy_breakdown} shows the number of requests seen per user/day under each of these four categories, separating the instances in which sensitive data was exposed from the total permission requests observed. Unlike Section \ref{sec:invisible}, we include ``visible'' permission requests (i.e., those occurring while the user was actively using the application or had other contextual information to indicate it was running).

Of the location permission checks, a majority were due to requests for location provider information (e.g., {\tt getBestProvider()} returns the best location provider based on application requirements), or checking WiFi state (e.g., {\tt getWifiState()} only reveals whether WiFi is enabled).  Only a portion of the requests actually exposed participants' locations (e.g., {\tt getLastKnownLocation()} or {\tt getScanResults()} exposed SSIDs of nearby WiFi networks). 


Although a majority of requests for the READ\_SMS permission exposed content in the SMS store (e.g., {\tt Query()} reads the contents of the SMS store), a considerable portion simply read information about the SMS store (e.g., {\tt renewMmsConnectivity()} resets an applications' connection to the MMS store). An exception to this is the use of SEND\_SMS, which resulted in the transmission of an SMS message every time the permission was requested.

Regarding browser history, both accessing visited URLs ({\tt getAllVisitedUrls()}) and reorganizing bookmark folders ({\tt addFolderToCurrent()}) result in the same permission being checked. However, the latter does not expose specific URLs to the invoking application.

Based on our analysis of the API methods that resulted in permission checks, we observed that sensitive data was only exposed to applications during half of these permission checks, on average. For instance, across both visible and invisible application requests, 5,111 of the 11,598 (44.3\%) permission checks involving the 12 permissions in Table \ref{tbl:perm-list} resulted in sensitive data being exposed to those applications (Table \ref{tbl:privacy_breakdown}).


While narrowing runtime permission requests down to only the cases in which sensitive resources are being accessed will greatly decrease the number of times users will be interrupted, the frequency with which these requests occur is still too great to reasonably prompt the user each time a request occurs. We also do not believe that the iOS model of only prompting on the first request is appropriate, because our data shows that in the vast majority of cases when protected resources are accessed, the user has no contextual cues to understand that it is occurring. Thus, a user may grant a request the first time an application asks, because it is appropriate in that instance, but then the user may be surprised to find that the application continues to access the resource in other contexts (e.g., when the user is not using the application). As a result, a more intelligent method is needed to determine when a given permission request is likely to be deemed appropriate by the user.

\section{User Expectations and Reactions}

To examine when users might want to be prompted about permission requests at runtime, our exit survey focused on participants' reactions to the 12 permissions in Table \ref{tbl:perm-list}, as well as limiting the number of requests shown to each participant based on our reservoir sampling algorithm, which was designed to ask participants about a diverse set of permission/application combinations. Thus, we collected participants' reactions to 673 permission requests ($\approx19$ per participant). Of these requests, 423 were accompanied by screenshots because participants were actively using their phones when the resources were accessed, whereas 250 permission requests were performed while device screens were off.\footnote{Our first 11 participants did not answer questions about permission requests occurring while not using their devices, and therefore the data only corresponds to our last 25 participants.} Of the former, 243 screenshots were taken while the requesting application was visible in the foreground, whereas 180 were taken while the application was invisible in the background.  In this section, we describe the situations in which permission requests defied users' expectations. We present explanations for why participants wanted to block certain permission requests, the factors influencing those decisions, and how expectations changed when devices were not actively in use.


\subsection{Reasons for Blocking}
When viewing screenshots of what they were doing when an application requested a permission, 30 participants (80\% of 36) stated that they would have preferred to block at least one request, whereas 6 stated a willingness to allow all requests, regardless of resource type or application. Across the entire study, participants wanted to block 35\% of these 423 permission requests. When we asked participants to explain their rationales for these decisions, two main themes emerged: the request did not---in their minds---pertain to application functionality or it involved information they were uncomfortable sharing.

\subsubsection{Relevance to Application Functionality}

When prompted for the reason behind blocking a permission request, 19 (53\% of 36) participants did not believe it was necessary for the application to perform its task. Of the 149 (35\% of 423) requests that participants would have preferred to block, 79 (53\%) were perceived as being irrelevant to the functionality of the application:

\begin{packed_item}
\item\textit{``It wasn't doing anything that needed my current location.''} (P1)
\item\textit{``I don't understand why this app would do anything with SMS.''} (P10)
\item\textit{``I don't know why my alarm clock needs to know where I am.''} (P8)
\end{packed_item}

Accordingly, functionality was the most common reason for wanting a permission request to proceed. Out of the 274 permissible requests, 195 (71\% of 274) were perceived as necessary for the core functionality of the application. Thirty-one (86\% of 36) participants mentioned perceived functional necessities as a reason for allowing at least one permission request to proceed:

\begin{packed_item}
\item\textit{``Because it's a weather app and it needs to know where you are to give you weather information.''}(P13)
\item\textit{``I think it needs to read the SMS to keep track of the chat conversation''}(P12)
\end{packed_item}

Beyond being necessary for core functionality, participants wanted 10\% (27 of 274) of requests to proceed because they offered convenience; 90\% of these requests were for location data, and the majority of those applications were published under the Weather, Social, and Travel \& Local categories in the Google Play store:

\begin{packed_item}
\item\textit{``It selects the closest stop to me so I don't have to scroll through the whole list.''} (P0)
\item\textit{``This app should read my current location. I'd like for it to, so I won't have to manually enter in my zip code / area.''} (P4)
\end{packed_item}

Thus, requests were allowed when they were expected: when participants rated the extent to which each request was expected on a 5-point Likert scale, allowable requests averaged 3.2, whereas blocked requests averaged 2.3 (lower is less expected).

\subsubsection{{Privacy Concerns}}

Participants also wanted to deny permission requests that involved data that they considered sensitive, regardless of whether they believed the application actually needed the data to function. Nineteen (53\% of 36) participants noted privacy as a concern while blocking a request, and of the 149 requests that participants wanted to block, 49 (32\% of 149) requests were blocked for this reason:

\begin{packed_item}
\item\textit{``SMS messages are quite personal.''} (P14)
\item\textit{``It is part of a personal conversation.''} (P11)
\item\textit{``Pictures could be very private and I wouldn't like for anybody to have access.''} (P16)
\end{packed_item}

Conversely, 24 participants (66\% of 36) wanted requests to proceed simply because they did not believe that the data involved was particularly sensitive; this reasoning accounted for 21\% of the 274 allowable requests:

\begin{packed_item}
\item\textit{``I'm ok with my location being recorded, no concerns.''} (P3)
\item\textit{``No personal info being shared.''} (P29)
\end{packed_item}

\subsection{Influential Factors}

Based on participants' responses to the 423 permission requests involving screenshots (i.e., requests occurring while they were actively using their phones), we quantitatively examined how various factors influenced their desire to block some of these requests.

\textbf{Effects of Identifying Permissions on Blocking}: In the exit survey, we asked participants to guess the permission an application was requesting, based on the screenshot of what they were doing at the time. The real answer was among four other incorrect answers. Of the 149 cases where participants wanted to block permission requests, they were only able to correctly state what permission was being requested 24\% of the time; whereas when wanting a request to proceed, they correctly identified the requested permission 44\% (of 274) of the time. However, Pearson's product-moment test\footnote{Both measures were normally distributed.} did not yield a statistically significant correlation (r=-0.171, p\textless0.317).



\textbf{Effects of Visibility on Expectations}: We were particularly interested in exploring if a permission request originating from a foreground application (i.e., running visibly to the user) was more expected than one from a background application. Of the 243 visible permission requests that we asked about in our exit survey, participants were able to correctly identify the requested permission 44\% of the time, and their average rating on our expectation scale was 3.4. On the other hand, participants were able to correctly identify the resources accessed by background applications only 29\% of the time (52 of 180), and their average rating on our expectation scale was 3.0. A Wilcoxon Signed-Rank test with continuity correction revealed a statistically significant difference in participants' expectations between these two groups (V= 441.5, p\textless0.001).


\textbf{Effects of Visibility on Blocking}: Participants were willing to block 71 (29\% of 243) permission requests originating from applications running in the foreground, whereas this increased by almost 50\% when the applications were running in the background invisible to them (43\% of 180). To examine whether invisible requests were more likely to be blocked, we calculated the percentage of denials for each participant, for both visible and invisible requests. A Wilcoxon Signed-Rank test with continuity correction revealed a statistically significant difference (V=58, p\textless0.001).



\textbf{Effects of Privacy Preferences on Blocking}: Participants completed a combination of Buchanan et al.'s Privacy Concerns Scale (PCS)~\cite{Buchanan2007} and Malhotra et al.'s Internet Users' Information Privacy Concerns (IUIPC) scale~\cite{Malhotra04}. A Spearman's rank test yielded no statistically significant correlation between their privacy preferences and their desire to block permission requests (\begin{math}\rho=0.156\end{math}, p\textless0.364).

\textbf{Effects of Expectations on Blocking}: We examined whether participants' expectations surrounding requests correlated with their desire to block them. For each participant, we calculated their average Likert scores for their expectations and the percentage of requests that they wanted to block.  Pearson's product-moment test showed a statistically significant correlation (r=-0.39, p\textless0.018). The negative correlation shows that participants were more likely to deny unexpected requests.

\subsection{User Inactivity and Resource Access}
In the second part of the exit survey, participants answered questions about 10 resource requests that occurred when the screen was off (not in use). Overall, they were more likely to expect resource requests to occur when using their devices ($\mu=3.26$ versus $\mu=2.66$). They also stated a willingness to block almost half of the permission requests (49.6\% of 250) when not in use, compared to a third of the requests that occurred when using their phones (35.2\% of 423). However, neither of these differences was statistically significant.


%
%

\section{Modeling Users' Decisions}
\label{sec:regressions}

We constructed several statistical models to examine whether users' desire to block certain permission requests could be predicted using the contextual data that we collected. If such a relationship exists, a classifier could determine when to prompt users about potentially unexpected permission requests. Thus, the response variable in our models is the user's choice to block the given permission request or not. Our predictive variables consisted of the information that might be available at runtime: permission type, requesting application, and visibility of that application. We constructed several mixed effects binary logistic regression models to account for both inter-subject and intra-subject correlations.

\subsection{Model Selection}
In our mixed effects models, permission types and visibility of the requesting application were fixed effects, because all possible values for each variable existed in our data set. Visibility had two values: visible (the user is interacting with the application or has other contextual cues to know that it is running) and invisible. Permission types were categorized based on Table \ref{tbl:privacy_breakdown}. The application name and the participant ID were included as random effects, because our survey data did not have an exhaustive list of all possible applications a user could run, and the participant has a non-systematic effect on the data.


Table \ref{tbl:models} shows two goodness-of-fit metrics: the Akaike Information Criterion (AIC) and Bayesian Information Criterion (BIC). Lower values for AIC and BIC represent better fit. Table \ref{tbl:models} shows the different parameters included in each model. We found no evidence of interaction effects and therefore did not include them. Visual inspection of residual plots of each model did not reveal obvious deviations from homoscedasticity or normality.

We initially included the phone's screen state as another variable. However, we found that creating two separate models based on the screen state resulted in better fit than using a single model that accounted for it as a fixed effect. When the screen was on, the model including application visibility and application name, while controlling for subject effects, offered the best fit. Here, fit improved once permission type was removed from the model, which shows that the decision to block a particular permission type changes based on contextual factors. When the screen is off, however, the effect of permission type was relatively stronger. Similarly, the strong subject effect in both models indicates that these decisions are highly nuanced from one user to the next. As a result, any classifier developed to automatically decide whether to block a permission at runtime (or prompt the user) will need to be tailored to that particular user's needs.

\begin{table}[t]
\centering
\small
\begin{tabular}{|l|l|l|l|}
\hline
\textbf{Predictors} & \textbf{AIC} & \textbf{BIC} &  \textbf{Screen State} \\ \hline
\hline
UserCode & 490.60 & 498.69 & Screen On \\ \hline
Application & 545.98 & 554.07 & Screen On \\ \hline
\begin{tabular}[c]{@{}l@{}}Application\\ UserCode\end{tabular} & 491.86 & 503.99 &  Screen On \\ \hline
\begin{tabular}[c]{@{}l@{}}Permission \\ Application\\ UserCode\end{tabular} & 494.69 & 527.05 &  Screen On \\ \hline
\textbf{\begin{tabular}[c]{@{}l@{}}Visibility \\Application\\ UserCode \end{tabular}} & \textbf{481.65} & \textbf{497.83}
 & \textbf{Screen On}\\ \hline
 \begin{tabular}[c]{@{}l@{}} Permission \\ Visibility \\ Application \\ UserCode \end{tabular} & 484.23 & 520.64 & Screen On \\ \hline
\hline

UserCode & 245.13 & 252.25 & Screen Off \\ \hline
Application & 349.38 & 356.50 &  Screen Off \\ \hline
\begin{tabular}[c]{@{}l@{}}Application\\ UserCode\end{tabular} & 238.84 & 249.52 &  Screen Off \\ \hline
\textbf{\begin{tabular}[c]{@{}l@{}}Permission  \\ Application \\ UserCode \end{tabular}} & \textbf{235.48} & \textbf{263.97} &  \textbf{Screen Off} \\ \hline

\end{tabular}
\begin{flushleft}
\caption{Goodness-of-fit metrics for various mixed effects logistic regression models on the exit survey data.}
\label{tbl:models}
\end{flushleft}
\end{table}

\subsection{Predicting User Reactions}

Using these two models, we built two classifiers to make decisions about whether to block any of the sensitive permission requests listed in Table \ref{tbl:privacy_breakdown}. These corresponded to 1.3M requests in our logs. We used our exit survey data as ground truth, randomly partitioning it into five groups, which we used for 5-fold cross-validation. We used four groups for training and one for evaluation. We iterated over the entire process 5 times and therefore report the averages over these 5 iterations.

We calculated the receiver operating characteristic (ROC) to capture the tradeoff between true-positive and false-positive rate. The quality of the classifier can be quantified with a single value by calculating the area under its ROC curve (AUC)~\cite{roc}. The closer the AUC gets to 1.0, the better the classifier is. The AUC for a random classifier is 0.5. When screens were on, the AUC was 0.7, which is 40\% better than the random baseline, making it a fair classifier. When screens were off, the AUC was 0.8, which is 60\% better than a random baseline, making it a good classifier. 

We ran the classifier on the log files we gathered from users and had it predict the portion of sensitive requests that users might want to block. When phone screens were on, the model predicted that participants would have blocked 35.29\% of sensitive requests, blocking an average of 645 requests daily. When screens were off, the classifier predicted 35.1\% of requests would have been blocked (1,143 requests per user/day).

\section{Discussion}

We observed that in one week of standard smartphone usage, 80\% of our participants deemed at least one permission request as being inappropriate. This violates Nissenbaum's notion of ``privacy as contextual integrity'' because applications were performing actions that defied users' expectations~\cite{nissenbaum2009privacy}. Felt et al.\ posited that users may be able to better understand why permission requests are needed if some of these requests are granted via runtime consent dialogs, rather than Android's current install-time notification approach~\cite{Felt2012b}. By granting permissions at runtime, users will have additional contextual information; based on what they were doing at the time that resources are requested, they may have a better idea of why those resources are being requested. We make two primary contributions that system designers can use to make more usable permissions systems: we show that runtime notifications need to be tailored to individual users' needs and that platforms need to account for whether an application's access to protected resources is obvious to the user.

Based on the frequency with which potential runtime permissions are requested (Section \ref{sec:apps}), it is infeasible to prompt users each and every time. Doing so would quickly overwhelm them and lead to habituation. Yet at the same time, drawing user attention to the situations in which they are likely to be concerned will lead to greater control and awareness. Thus, the challenge is in automatically inferring {\it when} users are likely to find a permission request unexpected, and then only prompting them in these cases. Based on our data, we observed that participants' desires to block particular permissions were heavily influenced by two main factors: their understanding of the relevance of a permission request to the functionality of the requesting application and their individual privacy concerns.

Our models in Section \ref{sec:regressions} showed that individual characteristics greatly explain the variance between what different users deem appropriate, in terms of access to protected resources. While responses to privacy scales failed to explain these differences, this was not a surprise, as the disconnect between stated privacy preferences and behaviors is well-documented (e.g.,~\cite{Acquisti05}). This means that in order to accurately model user preferences, the system will need to learn what a specific user deems inappropriate over time. Thus, a feedback loop is likely needed: when devices are ``new,'' users will be required to provide more input surrounding permission requests, and then based on their responses, they will see fewer requests in the future. While we documented the frequency with which protected resources were accessed, future work is needed to quantify the varying contexts in which it can occur. Specifically, habituation could be drastically reduced (i.e., fewer runtime prompts) if users are only asked for permission for unique combinations of application, resource, and context. Future work is needed to quantify the varying contexts, as they are likely to be much more complex than ``foreground vs. background application'' or ``screen on vs. screen off.''

Beyond individual subject characteristics (i.e., personal preferences), participants based their decisions to block certain permission requests on the specific applications making the requests and whether they had contextual cues to indicate that the applications were running (and therefore needed the data to function). Future systems could take these factors into account when deciding whether or not to draw user attention to a particular request. For example, when an application that a user is not actively using requests access to a protected resource, she should be shown a runtime prompt. If she decides to grant that request, then that decision is likely to hold when she is actively using that same application (and therefore a subsequent prompt may not be needed). At a minimum, platforms need to treat permission requests from background applications differently than those originating from foreground applications. Similarly, applications running in the background should use passive indicators to communicate when they are accessing particular resources. Platforms can also be designed to make decisions about whether or not access to resources should be granted based on whether contextual cues are present, or at its most basic, whether the device screen is even on.

Finally, we built our models and analyzed our data within the framework of what resources our participants {\it believed} were necessary for applications to correctly function. Obviously, their perceptions may have been incorrect in many cases, such that if they better understood why a particular resource was necessary, they may have been more permissive. Thus, it is incumbent on developers to adequately communicate why particular requests are necessary, as this greatly impacts user notions of contextual integrity. Yet, no mechanisms in Android exist for developers to do this as part of the permission-granting process. For example, one could imagine requiring metadata to be provided that explains how each requested resource will be used, and then automatically integrating this information into permission requests. Tan et al.\ examined a similar feature on iOS that allows developers to include free-form text in runtime permission dialogs and observed that users were significantly more likely to grant requests that included this text~\cite{Tan2014}. Thus, we believe that including succinct explanations in these requests would go a long way towards preserving contextual integrity by promoting greater transparency.

In conclusion, we believe this study was instructive in showing the circumstances in which Android permission requests are made under real-world usage. While prior work has already established that most mobile permissions systems are failing users, we believe our study can benefit system designers by demonstrating several ways in which contextual integrity can be improved, thereby empowering users to make better security decisions.

{\footnotesize \bibliographystyle{acm}
\bibliography{biblio}

\appendix

\section{Invisible requests}
\label{app:invisible}

Following list shows the set of applications that have requested most number of permissions while executing invisibly to the user and the most requested permission types by each respective application.

\begin{packed_item}
\item {\it Facebook App}--- ACCESS NETWORK STATE, ACCESS FINE LOCATION, ACCESS WIFI STATE ,WAKE LOCK, INTERNET
\item {\it Google Location}---WAKE LOCK, ACCESS FINE LOCATION, GET ACCOUNTS, ACCESS COARSE LOCATION, ACCESS WIFI STATE
\item {\it Facebook Messenger}---ACCESS NETWORK STATE, ACCESS WIFI STATE, WAKE LOCK, READ PHONE STATE, INTERNET
\item {\it Taptu DJ}---ACCESS NETWORK STATE, INTERNET, NFC
\item {\it Google Maps}---ACCESS NETWORK STATE, GET ACCOUNTS, WAKE LOCK, ACCESS FINE LOCATION, INTERNET
\item {\it Google (Gapps)}---WAKE LOCK, ACCESS FINE LOCATION, AUTHENTICATE ACCOUNTS, ACCESS NETWORK STATE, ACCESS WIFI STATE
\item {\it Fouraquare}---ACCESS WIFI STATE, WAKE LOCK, ACCESS FINE LOCATION, INTERNET, ACCESS COARSE LOCATION
\item {\it Yahoo Weather}---ACCESS FINE LOCATION, ACCESS NETWORK STATE, INTERNET, ACCESS WIFI STATE, WRITE SYNC SETTINGS
\item {\it Devexpert Weather}---ACCESS NETWORK STATE, INTERNET, ACCESS FINE LOCATION, ACCESS COARSE LOCATION
\item {\it Tile Game(Umoni)}---ACCESS NETWORK STATE, WAKE LOCK, INTERNET, ACCESS WIFI STATE, WRITE SETTINGS
\end{packed_item}

Following is the most frequently requested permission type by applications while running invisibly to the user and the applications who requested the respective permission type most.

\begin{packed_item}
\item {\it ACCESS\_NETWORK\_STATE}--- Facebook App, Google Maps, Facebook Messenger, Google (Gapps), Taptu - DJ
\item {\it WAKE\_LOCK}---Google (Location), Google (Gapps), Google (GMS), Facebook App, GTalk.
\item {\it ACCESS\_FINE\_LOCATION}---Google (Location), Google (Gapps), Facebook App, Yahoo Weather, Rhapsody (Music)
\item {\it GET\_ACCOUNTS}---Google (Location), Google (Gapps), Google (Login), Google (GM), Google (Vending)
\item {\it ACCESS\_WIFI\_STATE}---Google (Location), Google (Gapps), Facebook App, Foursqaure, Facebook Messenger
\item {\it UPDATE\_DEVICE\_STATS}---Google (SystemUI), Google (Location), Google (Gapps)
\item {\it ACCESS\_COARSE\_LOCATION}---Google (Location), Google (Gapps), Google (News), Facebook App, Google Maps
\item {\it AUTHENTICATE\_ACCOUNTS}---Google (Gapps), Google (Login), Twitter, Yahoo Mail, Google (GMS)
\item {\it READ\_SYNC\_SETTINGS}---Google (GM), Google ( GMS ), android.process.acore, Google (Email), Google (Gapps)
\item {\it INTERNET}---Google (Vending), Google (Gapps), Google (GM), Facebook App, Google (Location)
\end{packed_item}

\newpage

\section{Permission Type Breakdown}
\label{app:prem}

This table lists the most frequently used permissions during the study period.

{\center
\small
\begin{tabular}{|l|l|}
\hline
\textbf{Permission Type} & \textbf{Requests} \\ \hline
ACCESS\_NETWORK\_STATE & 41077 \\ \hline
WAKE\_LOCK & 27030 \\ \hline
ACCESS\_FINE\_LOCATION & 7400 \\ \hline
GET\_ACCOUNTS & 4387 \\ \hline
UPDATE\_DEVICE\_STATS & 2873 \\ \hline
ACCESS\_WIFI\_STATE & 2092 \\ \hline
ACCESS\_COARSE\_LOCATION & 1468 \\ \hline
AUTHENTICATE\_ACCOUNTS & 1335 \\ \hline
READ\_SYNC\_SETTINGS & 836 \\ \hline
VIBRATE & 740 \\ \hline
INTERNET & 739 \\ \hline
READ\_SMS & 611 \\ \hline
READ\_PHONE\_STATE & 345 \\ \hline
STATUS\_BAR & 290 \\ \hline
WRITE\_SYNC\_SETTINGS & 206 \\ \hline
CHANGE\_COMPONENT\_ENABLED\_STATE & 197 \\ \hline
CHANGE\_WIFI\_STATE & 168 \\ \hline
READ\_CALENDAR & 166 \\ \hline
ACCOUNT\_MANAGER & 134 \\ \hline
ACCESS\_ALL\_DOWNLOADS & 127 \\ \hline
READ\_EXTERNAL\_STORAGE & 126 \\ \hline
USE\_CREDENTIALS & 101 \\ \hline
READ\_LOGS & 94 \\ \hline
WRITE\_SMS & 62 \\ \hline
WRITE\_CALENDAR & 60 \\ \hline
Bluetooth & 60 \\ \hline
CONNECTIVITY\_INTERNAL & 58 \\ \hline
STATUS\_BAR\_SERVICE & 57 \\ \hline
READ\_SYNC\_STATS & 56 \\ \hline
NFC & 51 \\ \hline
WRITE\_SETTINGS & 50 \\ \hline
WRITE\_EXTERNAL\_STORAGE & 48 \\ \hline
PACKAGE\_USAGE\_STATS & 28 \\ \hline
SUBSCRIBED\_FEEDS\_READ & 24 \\ \hline
\end{tabular}
}

\newpage

\section{User Application Breakdown}
\label{app:app_brekdown}

This table shows the applications that most frequently requested access to protected resources during the study period.
{\center
\small
\begin{tabular}{|l|l|}
\hline
\textbf{Application Name} & \textbf{Requests} \\ \hline
facebook.katana & 40041 \\ \hline
google.process.location & 32426 \\ \hline
facebook.orca & 24702 \\ \hline
taptu.streams & 15188 \\ \hline
google.android.apps.maps & 6501 \\ \hline
google.process.gapps & 5340 \\ \hline
yahoo.mobile.client.android.weather & 5505 \\ \hline
tumblr & 4251 \\ \hline
king.farmheroessaga & 3862 \\ \hline
joelapenna.foursquared & 3729 \\ \hline
telenav.app.android.scout\_us & 3335 \\ \hline
devexpert.weather & 2909 \\ \hline
ch.bitspin.timely & 2549 \\ \hline
umonistudio.tile & 2478 \\ \hline
king.candycrushsaga & 2448 \\ \hline
android.systemui & 2376 \\ \hline
bambuna.podcastaddict & 2087 \\ \hline
contapps.android & 1662 \\ \hline
handcent.nextsms & 1543 \\ \hline
foursquare.robin & 1408 \\ \hline
qisiemoji.inputmethod & 1384 \\ \hline
devian.tubemate.home & 1296 \\ \hline
google.android.gm & 1180 \\ \hline
lookout & 1158 \\ \hline
aol.mobile.aim & 1125 \\ \hline
google.android.music:main & 1114 \\ \hline
rhapsody & 1074 \\ \hline
google.android.gms & 1021 \\ \hline
\begin{tabular}[c]{@{}l@{}}yahoo.mobile.client.android.\\fantasyfootball \end{tabular}
& 886 \\ \hline
jb.gosms & 755 \\ \hline
google.android.googlequicksearchbox & 711 \\ \hline
android.mms & 695 \\ \hline
ideashower.readitlater.pro & 672 \\ \hline
android.inputmethod.latin & 672 \\ \hline
google.android.gsf.login & 618 \\ \hline
pandora.android & 611 \\ \hline
google.android.apps.plus & 582 \\ \hline
andrewshu.android.reddit & 569 \\ \hline
tunein.player & 556 \\ \hline
airkast.KNBRAM & 546 \\ \hline
twitter.android & 523 \\ \hline
android.vending & 486 \\ \hline
yahoo.mobile.client.android.mail & 480 \\ \hline
sg.gumi.bravefrontier & 473 \\ \hline
yahoo.mobile.client.android.yahoo & 458 \\ \hline
nuance.swype.trial & 402 \\ \hline
viber.voip & 392 \\ \hline
zynga.words & 388 \\ \hline
touchtype.swiftkey & 377 \\ \hline
\end{tabular}
}

\newpage

\section{Distribution of Requests}
The two graphs shows the distribution of number of requests per day per user and the distribution of request for a given day. First graph shows the distribution of requests through out a given day averaged across the data set and second graph shows the distribution of number requests across days for each user.

\begin{figure}[h]
\centering
\begin{subfigure}[h]{0.5\textwidth}
\includegraphics[scale=0.45]{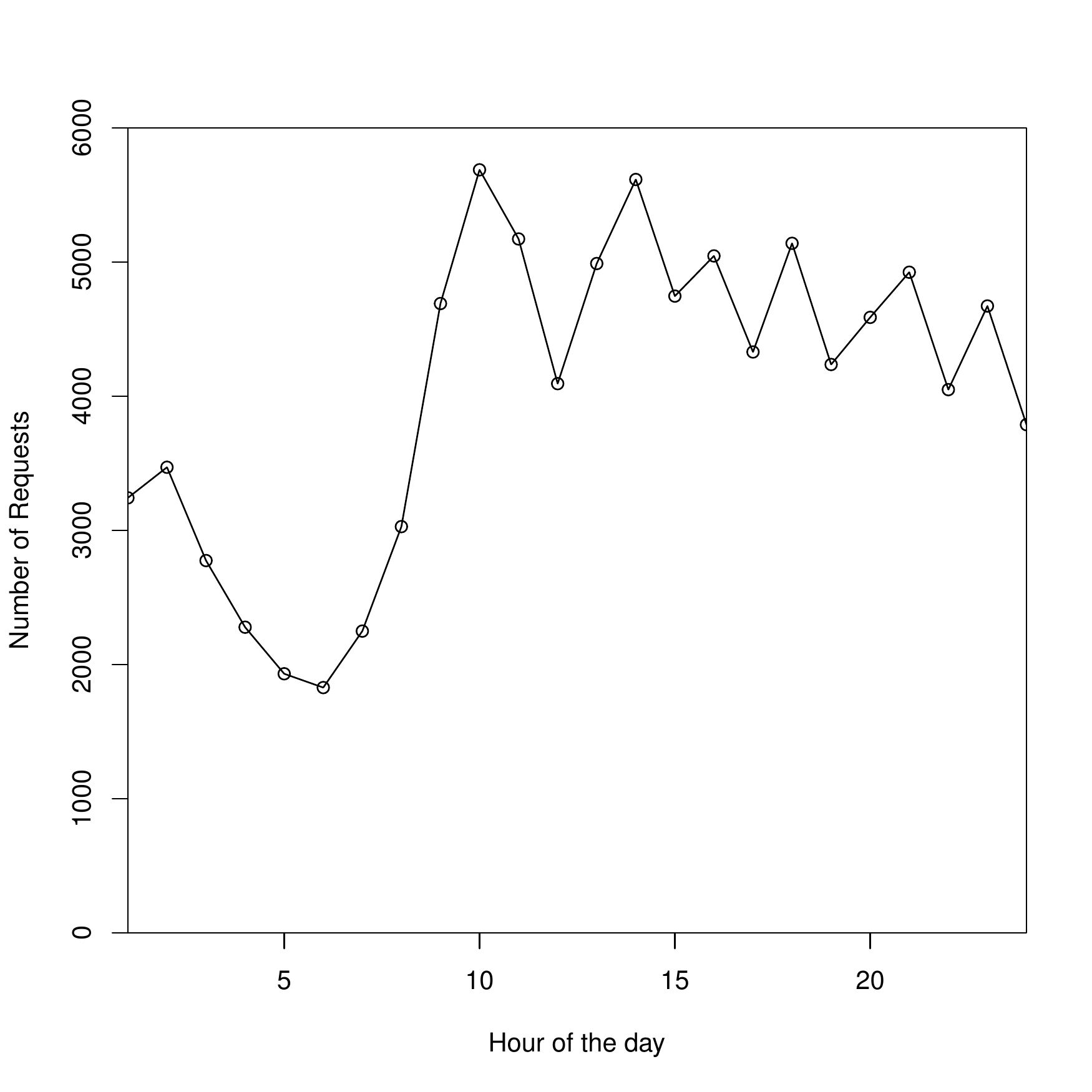}
\end{subfigure}
\quad
\begin{subfigure}[h]{0.5\textwidth}
\includegraphics[scale=0.45]{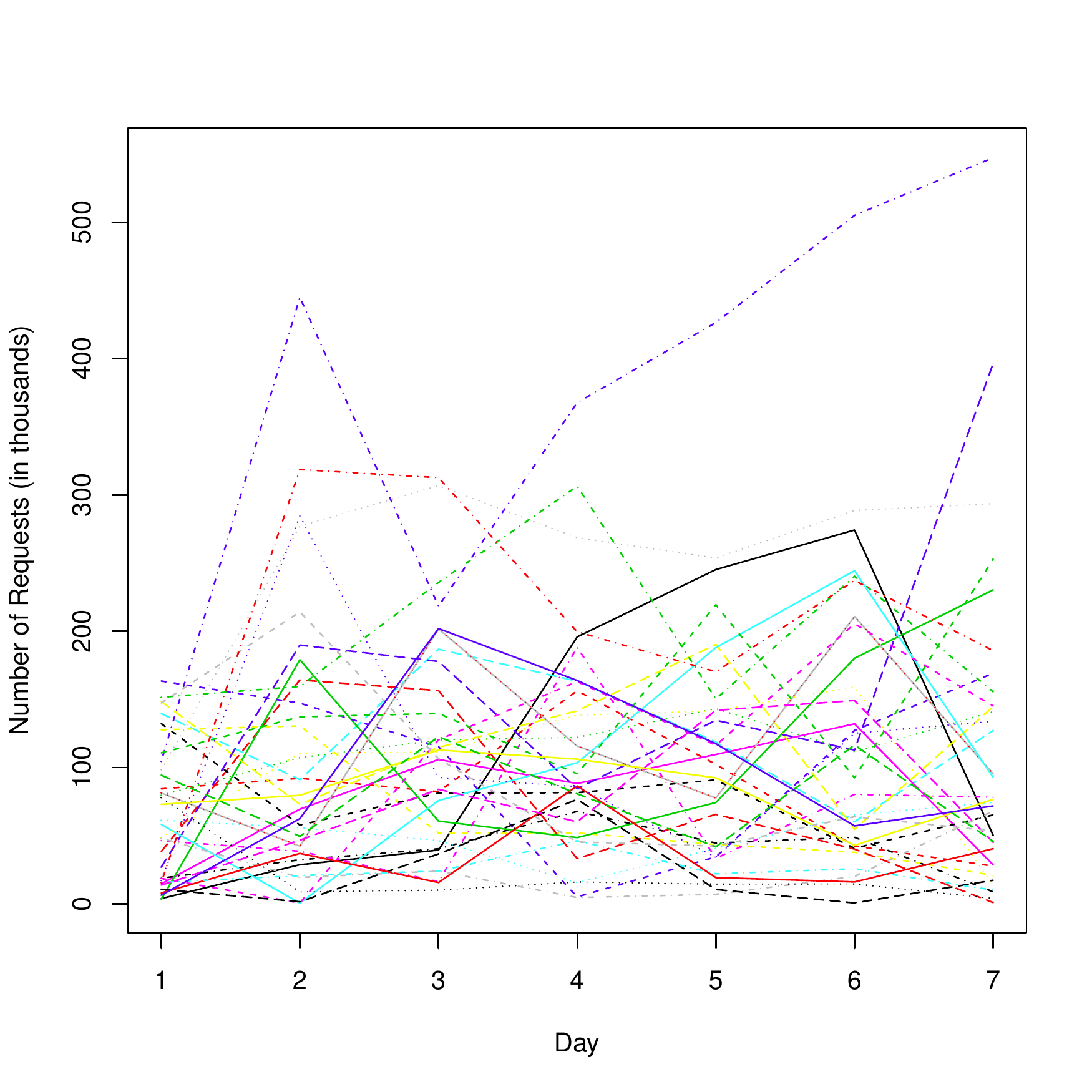}
\end{subfigure}
\begin{flushleft}
\end{flushleft}
\end{figure}
\end{document}